\documentclass[conference]{IEEEtran}
\usepackage[noadjust]{cite}

\ifCLASSINFOpdf
\else
\usepackage[dvips]{graphicx}
\DeclareGraphicsExtensions{.eps}
\fi
\usepackage{psfrag}

\usepackage[cmex10]{amsmath}
\usepackage{amssymb}
\usepackage[top=0.55in, bottom=0.7in, left=0.6in, right=0.6in]{geometry}

\def\uv{\mathbf{u}}
\def\cv{\mathbf{c}}
\def\xv{\mathbf{x}}
\def\hv{\mathbf{h}}
\def\yv{\mathbf{y}}
\def\wv{\mathbf{w}}
\def\zv{\mathbf{z}}
\def\muv{\boldsymbol{\mu}}
\def\Sgm{\boldsymbol{\Sigma}}
\def\I{\mathcal{I}}
\def\A{\mathcal{A}}
\def\N{\mathcal{N}}
\def\P{\mathcal{P}}
\def\D{\mathcal{D}}
\def\G{\mathcal{G}}
\def\Sd{\mathcal{S}_{\text{D}}}

\begin{document}
%
\title{Message-Passing Algorithms for Channel Estimation and Decoding Using Approximate Inference}

\author{
\IEEEauthorblockN{
Mihai-A. Badiu\IEEEauthorrefmark{1}\IEEEauthorrefmark{2},
Gunvor E. Kirkelund\IEEEauthorrefmark{1},
Carles Navarro Manch\'on\IEEEauthorrefmark{1},
Erwin Riegler\IEEEauthorrefmark{3} and
Bernard H. Fleury\IEEEauthorrefmark{1}}
\IEEEauthorblockA{
\IEEEauthorrefmark{1}
Aalborg University, Denmark
}
\IEEEauthorblockA{
\IEEEauthorrefmark{2}
Technical University of Cluj-Napoca, Romania
}
\IEEEauthorblockA{
\IEEEauthorrefmark{3}
Vienna University of Technology, Austria
}
}

\maketitle

\begin{abstract}

We design iterative receiver schemes for a generic communication system by treating channel estimation and information decoding as an inference problem in graphical models.
We introduce a recently proposed inference
framework that combines belief propagation (BP) and the mean field (MF)
approximation and includes these algorithms as special cases.
We also show that the expectation propagation and expectation maximization (EM) algorithms can be
embedded in the BP-MF framework with slight modifications.
By applying the considered inference algorithms to our probabilistic model, we derive four different
message-passing receiver schemes. Our numerical evaluation in a wireless scenario demonstrates that the receiver based on the BP-MF framework and its variant based on BP-EM yield
the best compromise between performance, computational complexity and numerical stability
among all candidate algorithms.

\end{abstract}


%
\IEEEpeerreviewmaketitle

\section{Introduction}

The design of advanced receiver algorithms is crucial to meet the stringent requirements of modern communication systems.
Motivated by the successful application of the ``turbo" principle in the decoding of channel codes, a large number of works have been devoted to the design of turbo receivers
(see~\cite{Tuchler11} and the references therein).
While in many of these works the receiver modules are individually designed and heuristically interconnected to exchange soft values, iterative receiver algorithms can be rigourously designed and better understood as instances of message-passing inference techniques (e.g., see~\cite{Boutros2002}).

In this context, variational Bayesian inference in probabilistic models~\cite{Wainwright2008} have proven to be a very useful tool to design receivers where tasks like channel estimation, detection and decoding are jointly derived.
Among the variational techniques, belief propagation (BP)~\cite{Kschischang2001,Loeliger2007} has found the most widespread use.
Originally applied to the decoding of channel codes, BP has been shown to be especially efficient in discrete probabilistic models. An alternative to BP is the mean field (MF) approximation and its message-passing counterpart, usually referred to as variational message-passing~\cite{Winn2005}.
MF inference has been successfully applied to continuous probabilistic models involving probability density functions (pdfs) belonging to an exponential family, in which BP suffers from numerical intractability.
Other notable examples of general-purpose inference techniques are expectation-maximization (EM)~\cite{Dempster1977} and expectation propagation (EP)~\cite{Minka2001}. EM is a special case of MF, where the approximate pdfs -- referred to as \emph{beliefs} -- are Dirac delta functions; EP can be seen as an approximation of BP where some beliefs are approximated by pdfs in a specific exponential family. Some attempts to find a unified framework encompassing all these techniques include the $\alpha$-divergence interpretation in~\cite{Minka05} and the region-based free energy approximations
in~\cite{Yedidia2005}. Following the latter approach, a novel hybrid message-passing inference framework combining BP and the MF approximation was recently proposed in~\cite{Riegler2012}.

In this paper, we investigate the design of receivers that perform joint channel estimation and data decoding in a generic communication system.
For this purpose, we capitalize on the combined inference framework~\cite{Riegler2012}, which provides some degree of freedom in the choice of the parts of the factor graph in which either BP or MF is applied. We show that this framework can be modified to naturally embed EP, EM and BP with Gaussian approximation of some messages.
Then, we apply these hybrid inference techniques to the underlying probabilistic model of the system and obtain four receiver algorithms, whose performance we assess by simulating a wireless system.

\emph{Notation}: we denote by $|\I|$ the cardinality of a finite set $\mathcal{I}$; the relative complement of $\{i\}$ in $\mathcal{I}$ is written as $\I\setminus i$; the set $\{ i \in \mathbb{N} \mid 1\leq i \leq n\}$ is denoted by $[1:n]$. Boldface lowercase and uppercase letters are used to represent vectors and matrices, respectively; superscripts ${(\cdot)}^{\operatorname{T}}$ and ${(\cdot)}^{\operatorname{H}}$ denote transposition and Hermitian transposition, respectively. The Hadamard product of two vectors is denoted by $\odot$.
For a vector $\xv = ( x_i \mid i \in \I )^{\operatorname{T}}$, we write $\xv_{\bar{i}} = ( x_j \mid j \in \I \setminus i )^{\operatorname{T}}$; for a matrix $\mathbf{A} \in \mathbb{C}^{m \times n}$, $\left[ \mathbf{A} \right]_{i,j}$ denotes its $(i,j)$th entry, $\left[ \mathbf{A} \right]_{\bar{i},\bar{j}}$ is the matrix $\mathbf{A}$ with the $i$th row and $j$th column deleted, $\left[ \mathbf{A} \right]_{\bar{i},j}$ denotes the column vector $( \left[ \mathbf{A} \right]_{k,j} \mid k \in [1:m] \setminus i )^{\operatorname{T}}$, and $\left[ \mathbf{A} \right]_{i,\bar{j}}$ is the row vector $( \left[ \mathbf{A} \right]_{i,k} \mid k \in [1:n] \setminus j )$.
The pdf of a multivariate complex Gaussian distribution with mean $\muv$ and covariance matrix $\Sgm$ is denoted by $\text{CN}(\cdot;\muv,\Sgm)$. We write $f(x) \propto g(x)$ when $f(x) = cg(x)$ for some positive constant $c$. We denote by $\G[\cdot]$ the approximation of the pdf in the argument with a Gaussian pdf with the same mean and covariance matrix.
The Dirac delta function is denoted by $\delta(\cdot)$.

\section{Message-Passing Inference Algorithms} \label{InferFrameworks}
We begin by concisely describing the unified message-passing
algorithm that combines the BP and MF approaches (refer
to~\cite{Riegler2012}). Then, we briefly show how
other widespread inference algorithms can be obtained as particular
instances or slight modifications of the unified framework.
%

Let $p(\zv)$ be an arbitrary pdf of a random vector $\zv \triangleq \left( z_i \mid i \in \I \right)^{\operatorname{T}}$ which factorizes as
\begin{equation}\label{eq:arbit_pdf}
p(\zv) = \prod_{a \in \A}f_a(\zv_a) = \prod_{a\in
\A_{\text{MF}}}f_a(\zv_a) \prod_{c\in \A_{\text{BP}}}f_c(\zv_c)
\end{equation}
where $\zv_a$ is the vector of all variables $z_i$ that are
arguments of the function $f_a$.
We have grouped the factors into two sets that partition $\A$: $\A_{\text{MF}}\cap \A_{\text{BP}}=\emptyset$ and $A_{\text{MF}}\cup \A_{\text{BP}}=\A$.
The factorization in~\eqref{eq:arbit_pdf} can be visualized in a factor graph~\cite{Kschischang2001} representation. We define $\N(a)\subseteq\I$ to be the set of indices of all variables $z_i$ that are arguments of function $f_a$; similarly, $\N(i)\subseteq\A$ denotes the set of indices of all functions $f_a$ that depend on $z_i$.
The parts of the graph that correspond to $\prod_{a\in \A_{\text{BP}}}f_a(\zv_a)$ and to $\prod_{a\in \A_{\text{MF}}}f_a(\zv_a)$ are referred to as ``BP part'' and ``MF part'', respectively.
We denote the variable nodes in the BP part by $\I_\text{BP}\triangleq \bigcup_{a\in\A_{\text{BP}}} \N(a)$ and those in the MF part by $\I_\text{MF}\triangleq \bigcup_{a\in\A_{\text{MF}}} \N(a)$.


The combined BP-MF inference algorithm approximates the marginals
$p(z_i)=\int p(\zv)\mathrm{d}\zv_{\bar i}$, $i\in\I$ by auxiliary
pdfs $b_i(z_i)$ called \emph{beliefs}. They are computed as \cite{Riegler2012}
\begin{equation}\label{eq:beliefBPMF}
    b_i(z_i) = \omega_i \prod\limits_{a \in \A_\text{BP} \cap \N(i)} m^{\text{BP}}_{a\to i}(z_i)
                   \prod\limits_{a \in \A_\text{MF} \cap \N(i)} m^{\text{MF}}_{a\to i}(z_i)
\end{equation}
with
\begin{equation} \label{eq:updateBPMF}
\begin{split}
m^{\text{BP}}_{a\to i}(z_i) &= \omega_a
\int \prod_{j\in \N(a)\setminus i} \operatorname{d} \! z_j \,n_{j\to a}(z_j) \,f_a(\zv_a),\\
&\forall\ a\in \A_\text{BP}, i\in\N(a) \\
m^{\text{MF}}_{a\to i}(z_i) &=
\exp \left( \int \prod_{j\in \N(a)\setminus i} \operatorname{d} \! z_j \, n_{j\to a}(z_j) \ln f_a(\zv_a) \right),\\
&\forall\ a\in \A_\text{MF}, i\in\N(a) \\
n_{i\to a}(z_i) &= \omega_i \prod\limits_{c \in \A_\text{BP} \cap \N(i)
\setminus a} m^{\text{BP}}_{c\to i}(z_i)
\prod\limits_{c \in \A_\text{MF} \cap \N(i)} m^{\text{MF}}_{c\to i}(z_i),\\
&\forall\ i\in\N(a), a\in\A,
\end{split}
\end{equation}
where $\omega_i$ and $\omega_a$ are constants that ensure normalized beliefs.


\emph{Belief propagation} is obtained as a particular case of BP-MF by setting $\A_\text{MF}=\emptyset$, since in this case the expressions in \eqref{eq:updateBPMF} reduce to the BP message computations.
Similarly, \emph{mean field} is an instance of BP-MF when $\A_\text{BP}=\emptyset$.

\emph{Expectation propagation} is very similar to BP, the main difference being that it constrains the beliefs of some variables to be members of a specific exponential family. Assuming Gaussian approximations of the beliefs, EP can also be integrated in the BP-MF framework by modifying the messages
\begin{equation} \label{eq:EP_m_msg}
    m^{\text{EP}}_{a \to i}(z_i) \propto \frac{1}{n_{i \to a}(z_i)}\, \G \left[ n_{i \to a}(z_i) \, m^{\text{BP}}_{a \to i}(z_i) \right],
\end{equation}
for all $i\in\mathcal{I}_{\text{EP}}\subseteq \I_{\text{BP}}$, $a \in \N(i)\cap\A_\text{BP}$. 


The \emph{expectation-maximization} algorithm is a special case of MF
when the beliefs of some variables are constrained to be Dirac delta functions~\cite{Riegler2012}. Again, we include this approximation in the BP-MF framework.
This leads to $n_{i\to a}(z_i) = \delta(z_i - \tilde z_i)$ for all $i \in
\mathcal{I}_\text{EM}\subseteq \I_{\text{MF}}$ and $a \in \N(i)\cap \A_{\text{MF}}$, where $\tilde z_i$ maximizes the unconstrained belief \eqref{eq:beliefBPMF}. We refer to this modified algorithm as BP-EM.
%

\section{Probabilistic System Model}
In this section, we present the signal model of our inference problem and its graphical representation. These will establish the baseline for the derivation of message-passing receivers.

We analyze a system consisting of one transmitter and one receiver.
A message represented by a vector $\uv = \left( u_{k} \mid k\in[1:K] \right)^{\operatorname{T}}\in\{0,1\}^K$ of information bits is conveyed by sending $N$ data and $M$ pilot channel symbols having the sets of indices $\D\subseteq{[1:M+N]}$ and $\P\subseteq[1:M+N]$, respectively, such that $\D\cup\P=[1:M+N]$ and $\D\cap\P=\emptyset$.
Specifically, vector $\uv$ is encoded and interleaved using a rate $R=K/(NL)$ channel code and a random interleaver into the vector $\cv = (\cv_{n}^{\operatorname{T}} \mid \cv_{n} \in \{0,1\}^L, n \in [1:N])^{\operatorname{T}}$ of length $NL$. For each $n \in [1:N]$, the subvector $\cv_{n} = (c_n^{(1)},\ldots,c_n^{(L)})^{\operatorname{T}}$ is mapped to a data symbol $x_{i_{n}} \in \Sd$ with $i_{n}\in \D$, where $\Sd$ is a discrete complex modulation alphabet of size $2^L$. Symbols $\xv_{\text{D}} = ( x_{i} \mid i\in \D )^{\operatorname{T}}$ are multiplexed with pilot symbols $\xv_{\text{P}} = ( x_{j} \mid j \in \P )^{\operatorname{T}}$, which are randomly selected from a QPSK modulation alphabet.
Finally, the aggregate vector of channel symbols $\xv = ( x_{i} \mid i\in \D \cup \P )^{\operatorname{T}}$ is sent through a channel with the following input-output relationship:
\begin{equation}\label{eq:channel_IO}
\yv=\hv\odot\xv+\wv.
\end{equation}
The vector $\yv = ( y_i \mid i\in [1:M+N] )^{\operatorname{T}}$ contains the received signal samples, $\hv = ( h_i \mid i\in [1:M+N] )^{\operatorname{T}}$ is the vector of channel coefficients, and $\wv = ( w_i \mid i\in [1:M+N] )^{\operatorname{T}}$ contains the samples of additive noise and has the pdf $p(\wv) = \text{CN} ( \wv;\mathbf{0},\gamma^{-1}\mathbf{I}_{M+N} )$ for some positive component precision $\gamma$. Note that \eqref{eq:channel_IO} can model any channel with a multiplicative effect that is not affected by inter-symbol interference, e.g., a time-varying frequency-flat channel or the equivalent channel in the frequency domain in a multicarrier system.

\begin{figure}[!t]
  \centering
  \psfrag{fh}[][]{\small $f_{\text{H}}$}
  \psfrag{h1}[][]{\small $h_{i_1}$}
  \psfrag{V}[][]{\small $\vdots$}
  \psfrag{hj}[][]{\small $h_j$}
  \psfrag{hn}[][]{\small $h_{i_N}$}
  \psfrag{fd1}[][]{\small $f_{\text{D}_{i_1}}$}
  \psfrag{fp}[][]{\small $f_{\text{P}_j}$}
  \psfrag{fdn}[][]{\small $f_{\text{D}_{i_N}}$}
  \psfrag{x1}[][]{\small $x_{i_1}$}
  \psfrag{xn}[][]{\small $x_{i_N}$}
  \psfrag{fm1}[][]{\small $f_{\text{M}_1}$}
  \psfrag{fmn}[][]{\small $f_{\text{M}_N}$}
  \psfrag{c1}[][]{\small $c_1^{(1)}$}
  \psfrag{c1L}[][]{\small $c_1^{(L)}$}
  \psfrag{cn1}[][]{\small $c_N^{(1)}$}
  \psfrag{cnL}[][]{\small $c_N^{(L)}$}
  \psfrag{fc}[][]{\small $f_{\text{C}}$}
  \psfrag{u1}[][]{\small $u_1$}
  \psfrag{uk}[][]{\small $u_K$}
  \psfrag{fu1}[][]{\small $f_{\text{U}_1}$}
  \psfrag{fuk}[][]{\small $f_{\text{U}_K}$}

  \includegraphics[width=0.95\columnwidth]{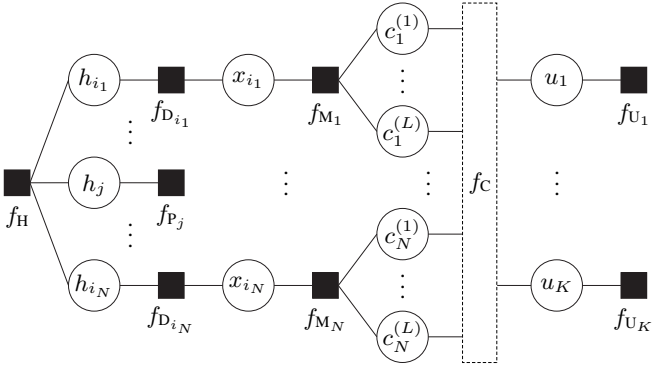}
  \caption{Factor graph representation of the pdf factorization in \eqref{eq:joint_pdf_f} with $i_1,\ldots,i_N \in \D$ and $j \in \P$.}
  \label{fig:FacGraph}
\end{figure}

Based on the above signal model, we can state the probabilistic model which captures the dependencies between the system variables. The pdf of the collection of observed and unknown variables factorizes as
\begin{align}
&p(\yv,\hv,\xv_{\text{D}},\cv,\uv) =  f_{\text{H}}(\hv) \prod_{i\in\D} f_{\text{D}_i}(h_i,x_i) \prod_{j\in\P} f_{\text{P}_j}(h_j) \nonumber\\
& \qquad \times \prod_{n\in[1:N]} f_{\text{M}_n} (x_{i_n},\cv_n) \, f_{\text{C}}(\cv,\uv) \prod_{k\in[1:K]}f_{\text{U}_k}(u_k), \label{eq:joint_pdf_f}
\end{align}
where $f_{\text{D}_i}(h_i,x_i) \triangleq p(y_i|h_i,x_i)$ and $f_{\text{P}_j}(h_j)\triangleq p(y_j|h_j)$ incorporate the observations in $\yv$ and are given by
\begin{align}
    f_{\text{D}_i}(h_i,x_i)
    &= \text{CN} \left( h_i x_i;y_i,\gamma^{-1} \right),\quad \forall i\in \D, \label{eq:ObsFacD} \\
    f_{\text{P}_j}(h_j)
    &= \text{CN} \left( h_j x_j;y_j,\gamma^{-1} \right),\quad \forall j\in \P, \label{eq:ObsFacP}
\end{align}
$f_{\text{H}}(\hv) \triangleq p(\hv)$ is the prior pdf of the vector of channel coefficients for which we set
\begin{equation}\label{eq:chan_prior}
    f_{\text{H}}(\hv) = \text{CN} \left( \hv;\muv_{\hv}^{\text{p}},\Sgm_{\hv}^{\text{p}} \right),
\end{equation}
$f_{\text{M}_n}\left( x_{i_n},\cv_n \right) \triangleq p \left( x_{i_n}|\cv_n \right)$ stand for the modulation mapping,
$f_{\text{C}}(\cv,\uv) \triangleq p(\cv|\uv)$ accounts for the coding and interleaving operations
and $f_{\text{U}_k}(u_k) \triangleq p(u_k)$ is the prior pmf of the $k$th
information bit. To obtain \eqref{eq:joint_pdf_f}, we used the fact that $\yv$ is conditionally independent of $\cv$ and $\uv$ given $\xv_{\text{D}}$, $\hv$ is independent of $\xv_{\text{D}}$, $\cv$ and $\uv$, the noise samples $w_i$ are i.i.d., and each data symbol $x_{i_n}$ is conditionally independent of all the other symbols given $\cv_n$. The factorization in \eqref{eq:joint_pdf_f} can be
visualized in the factor graph depicted in Fig.~\ref{fig:FacGraph}. The graph of the code and interleaver is not explicitly given, its structure being captured by $f_{\text{C}}$.

\section{Message-Passing Receiver Schemes} \label{ReceiverAlgorithms}

In this section, we derive iterative receiver schemes by applying different inference algorithms to the factor graph in Fig.~\ref{fig:FacGraph}. The receiver has to infer the beliefs of the information bits using the observed vector $\yv$ and prior knowledge, i.e., the pilot symbols and their set of indices $\P$, the noise precision $\gamma$, the channel statistics in \eqref{eq:chan_prior}, the modulation mapping and the structure of the channel code and interleaver.

We set $\A$ and $\I$ (defined in Section~\ref{InferFrameworks} for a general probabilistic model) to be the sets of all factors and variables, respectively, contained in our probabilistic model.
Next, we show that the BP algorithm resulting from setting $\A_\text{MF}=\emptyset$ yields messages of an intractable complexity. Assume that by running BP in the part of the graph containing the modulation and code constraints we obtain the messages
\begin{equation} \label{eq:m_fM_x}
    m^{\text{BP}}_{f_{\text{M}_n} \to x_{i_n}}(x_{i_n}) \propto \sum_{s \in \Sd} \beta_{i_n}(s)\delta(x_{i_n}-s),
\end{equation}
with $i_n \in \D, \forall n \in [1:N]$, where $\beta_{i_n}(s)$ represent extrinsic information on symbol $x_{i_n}$.
These messages are further passed as $n_{x_{i_n} \to f_{\text{D}_{i_n}}}(x_{i_n}) = m^{\text{BP}}_{f_{\text{M}_n} \to x_{i_n}}(x_{i_n})$.
Then, for each $i \in \D$, compute the message
\begin{align}
    m^{\text{BP}}_{f_{\text{D}_{i}} \to h_{i}}(h_{i}) & \propto \int f_{\text{D}_{i}}(h_i,x_i) \, n_{x_{i} \to f_{\text{D}_{i}}}(x_{i}) \operatorname{d} \! x_i \nonumber \\
    & \propto \sum_{s \in \Sd} \frac{\beta_i(s)}{|s|^2} \, \text{CN} \left( h_i;\frac{y_i s^\ast}{|s|^2},\frac{1}{\gamma |s|^2} \right), \label{eq:m_fD_h_BP}
\end{align}
while for all $j \in \P$ set
\begin{equation}\label{eq:m_fP_h}
    m^{\text{BP}}_{f_{\text{P}_j} \to h_j}(h_j) \propto f_{\text{P}_j}(h_j) \propto \text{CN} \left( h_j;\frac{y_j x_j^\ast}{|x_j|^2},\frac{1}{\gamma |x_j|^2} \right).
\end{equation}
Note that the message in~\eqref{eq:m_fD_h_BP} is proportional to a mixture of Gaussian pdfs with $|\Sd|=2^L$ components.
Then, after setting $n_{h_i \to f_{\text{H}}}(h_i) = m^{\text{BP}}_{f_{\text{D}_i} \to h_i}(h_i)$ for all $i \in \D$ and $n_{h_i \to f_{\text{H}}}(h_i) = m^{\text{BP}}_{f_{\text{P}_i} \to h_i}(h_i)$ for all $i \in \P$,
the message from $f_{\text{H}}$ to $h_i$ reads
\begin{equation}\label{eq:m_fH_h_intract}
    m^{\text{BP}}_{f_{\text{H}} \to h_{i}}(h_{i}) \propto \int f_{\text{H}}(\hv) \prod_{j \in (D \cup P) \setminus i}{n_{h_j \to f_{\text{H}}}(h_j) \operatorname{d} \! h_j}.
\end{equation}
Using \eqref{eq:chan_prior}, \eqref{eq:m_fD_h_BP} and \eqref{eq:m_fP_h}, the message in~\eqref{eq:m_fH_h_intract} becomes a Gaussian mixture with $2^{L(N-1)}$ and $2^{LN}$ components for $i \in \D$ and $i\in\P$, respectively.
Clearly, the computation of such messages is intractable and one has to use approximations.

\subsection{Algorithm based on BP combined with Gaussian approximation} \label{BP-GA}

Since the intractability of the messages occurs due to the Gaussian mixture in \eqref{eq:m_fD_h_BP}, we approximate those messages as proposed in~\cite{ShWoHoAu10}, i.e., for each $i \in \D$ we set
\begin{equation}\label{eq:BP_Gapprox}
m^{\text{BP-GA}}_{f_{\text{D}_i} \to h_i}(h_i) \propto \G \left[ m^{\text{BP}}_{f_{\text{D}_i} \to h_i}(h_i) \right] = \text{CN} \left( h_i;\mu_{h_i,\text{o}},\sigma^2_{h_i,\text{o}} \right)
\end{equation}
with
\begin{equation}\label{eq:Gapprox_mean_var}
\begin{split}
    \mu_{h_i,\text{o}} & = \sum_{s \in \Sd}{\alpha_i(s) \frac{y_i s^\ast}{|s|^2}},\\
    \sigma^2_{h_i,\text{o}} & = \sum_{s \in \Sd}{\alpha_i(s) \left( \frac{|y_i|^2}{|s|^2} + \frac{1}{\gamma|s|^2}\right)} - \left| \mu_{h_i,\text{o}} \right|^2.
\end{split}
\end{equation}
In~\eqref{eq:Gapprox_mean_var}, we have defined the normalized amplitudes of the Gaussian mixture $\alpha_i(s) = \beta_i(s)/(\kappa_i|s|^2)$, where the constant
$\kappa_i$ ensures $\sum_{s \in \Sd}\alpha_i(s) = 1$.
We also denote the mean and variance of the pdf in~\eqref{eq:m_fP_h} by $\mu_{h_j,\text{o}}$ and $\sigma^2_{h_j,\text{o}}$, $j \in \P$, and we define the vector $\muv_{\hv}^\text{o} = \left( \mu_{h_i,\text{o}} \mid i \in [1:M+N] \right)^{\operatorname{T}}$ and the matrix $\Sgm^{\text{o}}_{\hv}$ with entries $[\Sgm^{\text{o}}_{\hv}]_{i,j} = \sigma^2_{h_i,\text{o}}$ if $i=j$ and zero otherwise, for all $i,j \in [1:M+N]$.

Now, using \eqref{eq:chan_prior} and \eqref{eq:BP_Gapprox}, the message in~\eqref{eq:m_fH_h_intract} becomes
\begin{align} \label{eq:m_fH_h_BPapprox}
m^{\text{BP}}_{f_{\text{H}} \to h_{i}}(h_{i}) & \propto \int \text{CN} \left( \hv;\muv_{\hv}^{\text{p}},\Sgm^{\text{p}}_{\hv} \right) \text{CN} \left( \hv_{\bar{i}};\muv^{\text{o}}_{\hv_{\bar{i}}},[\Sgm^{\text{o}}_{\hv}]_{\bar{i},\bar{i}} \right) \operatorname{d} \! \hv_{\bar{i}} \nonumber \\
& \propto \text{CN} \left( h_i; \mu_{h_i,\text{c}}, \sigma^2_{h_i,\text{c}} \right),
\end{align}
with
\begin{equation} \label{eq:BP_hc_param}
\begin{split}
\mu_{h_i,\text{c}} &= \mu^p_{h_i} + [\Sgm^{\text{p}}_{\hv}]_{i,\bar{i}} \left( [\Sgm^{\text{o}}_{\hv}]_{\bar{i},\bar{i}} + [\Sgm^{\text{p}}_{\hv}]_{\bar{i},\bar{i}} \right)^{-1} (\muv^{\text{o}}_{\hv_{\bar{i}}} - \muv^{\text{p}}_{\hv_{\bar{i}}}), \\
\sigma^2_{h_i,\text{c}} &= [\Sgm^{\text{p}}_{\hv}]_{i,i} - [\Sgm^{\text{p}}_{\hv}]_{i,\bar{i}} \left( [\Sgm^{\text{o}}_{\hv}]_{\bar{i},\bar{i}} + [\Sgm^{\text{p}}_{\hv}]_{\bar{i},\bar{i}} \right)^{-1} [\Sgm^{\text{p}}_{\hv}]_{\bar{i},i}.
\end{split}
\end{equation}
These messages are further passed as extrinsic values, i.e., $n_{h_i \to f_{\text{D}_i \text{ or P}_i}}(h_i) = m^{\text{BP}}_{f_{\text{H}} \to h_{i}}(h_{i})$. For each $i \in \D$, the following message is then computed:
\begin{align*}
    m^{\text{BP}}_{f_{\text{D}_{i}} \to x_i}(x_i) & \propto \int f_{\text{D}_{i}}(h_i,x_i) \, n_{h_i \to f_{\text{D}_{i}}}(h_i) \operatorname{d} \! h_i \nonumber \\
    & \propto \frac{1}{\gamma^{-1} + \sigma^2_{h_i,\text{c}} |x_i|^2 } \exp \left( - \frac{ \left| y_i - \mu_{h_i,\text{c}}x_i \right|}{\gamma^{-1} + \sigma^2_{h_i,\text{c}} |x_i|^2 } \right).
\end{align*}
After passing the extrinsic messages $n_{x_{i_n} \to f_{\text{M}_n}}(x_{i_n}) = m^{\text{BP}}_{f_{\text{D}_{i_n}} \to x_{i_n}}(x_{i_n})$, $i_n \in \D$, $n \in [1:N]$, we apply the BP update rule to compute the probabilities of the coded and interleaved bits (which is equivalent to MAP demapping), followed by BP decoding to obtain the beliefs of the information bits.
\subsection{Algorithm based on expectation propagation}
We set $\A_\text{MF}=\emptyset$ and $\I_\text{EP} = \{ h_i \mid i \in \D \}$. The message $m^{\text{EP}}_{f_{\text{D}_{i}} \to h_{i}}(h_{i})$ computed with~\eqref{eq:EP_m_msg} is proportional to a Gaussian pdf; consequently, the EP rule for $m^{\text{EP}}_{f_{\text{H}} \to h_{i}}(h_{i})$ reduces to the BP rule and outputs a Gaussian pdf as in \eqref{eq:m_fH_h_BPapprox}, since the operator $\G[\cdot]$ is an identity operator for Gaussian arguments.

Specifically, using \eqref{eq:updateBPMF}, \eqref{eq:EP_m_msg}, and then \eqref{eq:m_fD_h_BP}, \eqref{eq:m_fH_h_BPapprox}, we have
\begin{equation*}
b_{h_i}(h_i) = \G \left[ n_{h_i \to f_{\text{D}_{i}}}(h_i) m^{\text{BP}}_{f_{\text{D}_{i}} \to h_{i}}(h_{i})  \right] = \text{CN} \left( h_i; \mu_{h_i}, \sigma^2_{h_i}\right),
\end{equation*}
for each $i \in \D$, where
\begin{equation*}
\begin{split}
\mu_{h_i} &= \sum_{s \in \Sd} \phi_i(s) \frac{\sigma^{-2}_{h_i,\text{c}} \, \mu_{h_i,\text{c}} + \gamma y_i s^\ast}{\sigma^{-2}_{h_i,\text{c}} + \gamma|s|^2}, \\
\sigma^2_{h_i} &=  \sum_{s \in \Sd} \phi_i(s) \frac{\left|\sigma^{-2}_{h_i,\text{c}} \, \mu_{h_i,\text{c}} + \gamma y_i s^\ast\right|^2 + \sigma^{-2}_{h_i,\text{c}} + \gamma|s|^2}{\left(\sigma^{-2}_{h_i,\text{c}} + \gamma|s|^2 \right)^2} - \left| \mu_{h_i} \right|^2
\end{split}
\end{equation*}
with
\begin{equation*}
\phi_i(s) \triangleq \frac{\beta_i(s)\, \text{CN} \left( y_i; \mu_{h_i,\text{c}} s, \gamma^{-1} + \sigma^{-2}_{h_i,\text{c}} |s|^2 \right)}{\sum_{s \in \Sd} \beta_i(s)\, \text{CN} \left( y_i; \mu_{h_i,\text{c}} s, \gamma^{-1} + \sigma^{-2}_{h_i,\text{c}} |s|^2 \right)}
\end{equation*}
and $\mu_{h_i,\text{c}}$, $\sigma^2_{h_i,\text{c}}$ as in \eqref{eq:BP_hc_param}.
Using \eqref{eq:EP_m_msg} again, we obtain
\begin{equation*}\label{eq:EP_fD_h}
m^{\text{EP}}_{f_{\text{D}_i} \to h_i}(h_i) \propto \frac{\text{CN} \left( h_i;\mu_{h_i},\sigma^2_{h_i} \right)}{\text{CN} \left( h_i;\mu_{h_i,\text{c}},\sigma^2_{h_i,\text{c}} \right)}
\propto \text{CN} \left( h_i;\mu_{h_i,\text{o}},\sigma^2_{h_i,\text{o}} \right),
\end{equation*}
with
\begin{equation}\label{eq:EP_mean_var}
\begin{split}
\sigma^{-2}_{h_i,\text{o}} & = \sigma^{-2}_{h_i} - \sigma^{-2}_{h_i,\text{c}}, \\
\mu_{h_i,\text{o}} & = \sigma^2_{h_i,\text{o}} \left( \sigma^{-2}_{h_i}\mu_{h_i} - \sigma^{-2}_{h_i,\text{c}}\mu_{h_i,\text{c}} \right).
\end{split}
\end{equation}
Unlike \eqref{eq:Gapprox_mean_var} in BP with Gaussian approximation, the values of $\mu_{h_i,\text{o}}$ and $\sigma^2_{h_i,\text{o}}$, $i \in \D$, computed with \eqref{eq:EP_mean_var} depend on all $\mu_{h_j,\text{o}}$ and $\sigma^2_{h_j,\text{o}}$, $j \in \D$, $j\neq i$, through \eqref{eq:BP_hc_param}.
The parameters of $m^{\text{EP}}_{f_{\text{H}} \to h_{i}}(h_{i})$ are updated using \eqref{eq:BP_hc_param} but with $\mu_{h_i,\text{o}}$ and $\sigma^2_{h_i,\text{o}}$ computed as above.
Note that all messages that depend on the channel coefficients need to be updated in a sequential manner.
The rest of the messages are computed as in Section \ref{BP-GA}.

\subsection{Algorithm based on the combined BP-MF framework}
The factor graph is split into the MF and BP parts by setting $\A_\text{MF} = \{f_{\text{D}_i}\mid i\in \D \}$ and $\A_\text{BP} = \A \setminus \A_\text{MF}$. Such a splitting yields tractable and simple messages, takes advantage of the fact that BP works well with hard constraints and best exploits the correlation between the channel coefficients for the graphical representation in Fig.~\ref{fig:FacGraph}\footnote{Alternatively, the same level of exploitation of the correlation is obtained by representing the channel variables as a single vector variable $\hv$ and ``moving'' factor node $f_\text{H}$ to the MF part~\cite{Riegler2012}.}.

Assuming we have obtained the messages $n_{x_i \to f_{\text{D}_i}}(x_i)$ (their expression will be given later), we can compute
\begin{align*}
    m^{\text{MF}}_{f_{\text{D}_i} \to h_i}(h_i) &\propto \exp \left(
        \int n_{x_i \to f_{\text{D}_i}}(x_i) \ln f_{\text{D}_i}(h_i,x_i) \operatorname{d} \! x_i
    \right) \\
    &\propto \text{CN} \left( h_i;\mu_{h_i,\text{o}},\sigma^2_{h_i,\text{o}} \right),
\end{align*}
where
\begin{equation*}
    \mu_{h_i,\text{o}} = \frac{y_i \mu^{\ast}_{x_i}}{\sigma^{2}_{x_i} + |\mu_{x_i}|^2},\quad
    \sigma^2_{h_i,\text{o}} = \frac{1}{\gamma \left( \sigma^{2}_{x_i} + |\mu_{x_i}|^2 \right)},
\end{equation*}
with the definition $\mu_{x_i}\triangleq \int n_{x_i \to f_{\text{D}_i}}(x_i) \, x_{i} \operatorname{d} \! x_i$ and $\sigma^2_{x_i}\triangleq \int n_{x_i \to f_{\text{D}_i}}(x_i) |x_{i}-\mu_{x_{i}}|^2 \operatorname{d} \! x_i$.

The messages $n_{h_i \to f_{\text{H}}}(h_i) = m^{\text{MF}}_{f_{\text{D}_i} \to h_i}(h_i)$ are sent to the BP part and hence are extrinsic values. When computing $m^{\text{BP}}_{f_{\text{H}} \to h_{i}}(h_{i})$ we get the same expression as \eqref{eq:m_fH_h_BPapprox}, with the parameters \eqref{eq:BP_hc_param}.
Unlike in the previous algorithms, the following messages are beliefs, i.e., \emph{a posteriori} probabilities (APP):
\begin{align*}
    n_{h_i \to f_{\text{D}_i}}(h_i) &= \omega_{h_i}\,m^{\text{BP}}_{f_{\text{H}} \to h_{i}}(h_{i}) \,m^{\text{MF}}_{f_{\text{D}_i} \to h_i}(h_i) \\
&= \text{CN} \left( h_i;\mu_{h_i},\sigma^2_{h_i} \right), \quad \forall i \in \D,
\end{align*}
with
\begin{align}
\mu_{h_i} &= \left( \sigma^{-2}_{h_i,\text{o}} + \sigma^{-2}_{h_i,\text{c}} \right)^{-1} \left( \sigma^{-2}_{h_i,\text{o}}\mu_{h_i,\text{o}} + \sigma^{-2}_{h_i,\text{c}}\mu_{h_i,\text{c}} \right), \label{eq:muH_BPMF} \\
\sigma^{-2}_{h_i} &= \sigma^{-2}_{h_i,\text{o}} + \sigma^{-2}_{h_i,\text{c}}. \nonumber
\end{align}
Then, for all $i \in \D$, we compute
\begin{align}
    &m^{\text{MF}}_{f_{\text{D}_i} \to x_i}(x_i) \propto \exp \left(
        \int n_{h_i \to f_{\text{D}_i}}(h_i) \ln f_{\text{D}_i}(h_i,x_i) \operatorname{d} \! h_i
    \right) \nonumber \\
    &\qquad \qquad \, \propto \text{CN} \left( x_i; \frac{y_i\mu_{h_i}^\ast}{\sigma_{h_i}^2+|\mu_{h_i}|^2},\frac{1} {\gamma (\sigma_{h_i}^2+|\mu_{h_i}|^2)} \right) \label{eq:m_fD_x_MF}
\end{align}
and we pass $n_{x_{i_n}\to f_{\text{M}_n}}(x_{i_n}) = m^{\text{MF}}_{f_{\text{D}_{i_n}} \to x_{i_n}}(x_{i_n})$ to the modulation and coding part of the graph as extrinsic values, for all $n \in [1:N]$. After running BP, we obtain \eqref{eq:m_fM_x} and then pass the following APP values back to the MF part:
\begin{align*}
    n_{x_{i_n} \to f_{\text{D}_{i_n}}}(x_{i_n}) &= \omega_{x_{i_n}}\,m^{\text{BP}}_{f_{\text{M}_n} \to x_{i_n}}(x_{i_n}) \,m^{\text{MF}}_{f_{\text{D}_{i_n}} \to x_{i_n}}(x_{i_n}).
\end{align*}
\subsection{Algorithm based on BP-EM}
We now apply EM for channel estimation, so we constrain $b_{h_i}(h_i)$ from the previous BP-MF scheme to be Dirac delta functions. The resulting messages are the same as in the previous subsection, except for $n_{h_i \to f_{\text{D}_i}}(h_i) = \delta(h_i - \mu_{h_i})$ with $\mu_{h_i}$ computed as in \eqref{eq:muH_BPMF}. Note that this algorithm uses only point estimates of the channel weights; however, its complexity is basically still the same, since the computation of \eqref{eq:muH_BPMF} actually includes the computation of the corresponding variance.

\subsection{Scheduling of message computations}

All algorithms employ the same message-passing scheduling: they start by sending messages $m_{f_{\text{P}_j} \to h_j}(h_j)$ corresponding to pilots and by initializing $m_{f_{\text{D}_i} \to h_i}(h_i) \propto \text{CN}\left(h_i;0,\infty \right)$; messages (computed according to the corresponding algorithm) are passed on up to the information bit variables -- this completes the first iteration; each following iteration consists in passing messages up to the channel prior factor node and back; messages are passed back and forth until a predefined number of iterations is reached. All algorithms end by taking hard decisions on the beliefs of the information bits.

\section{Simulation Results}
We consider a wireless OFDM system with the parameters given in Table~\ref{tab:sim_param}, and we evaluate by means of Monte Carlo simulations the bit error rate (BER) performance of the receiver
algorithms derived in Section \ref{ReceiverAlgorithms}. We employ as a reference a scheme which has perfect channel state information (CSI), i.e., it has prior knowledge of the vector of channel coefficients $\hv$.

We encountered numerical problems with the EP-based scheme due to the instability of EP in general, so we used the heuristic approach~\cite{Minka05} to damp the updates of the beliefs $b_{h_i}$ with a step-size $\epsilon=0.5$. Also, the EP-based scheme has higher computational complexity than the others due to its message definition -- it requires multiplication of a Gaussian pdf with a mixture of Gaussian pdfs, the approximation $\G[\cdot]$ and division of Gaussian pdfs -- and to the sequentiality of the message updates for the channel coefficients\footnote{For the other receiver schemes, it can be shown that the parameters of all messages $m^{\text{BP}}_{f_{\text{H}} \to h_{i}}(h_{i})$ with $i\in\D\cup\P$ can be computed jointly and with a lower complexity.}.

\begin{figure}[!t]
\centering
\includegraphics[width=0.75\columnwidth]{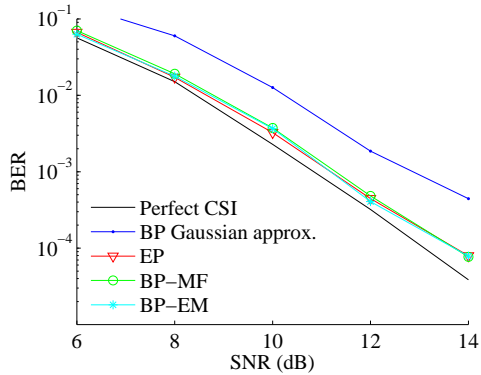}
\caption{BER vs. SNR performance of the receiver algorithms for a number of pilot symbols $M=10$, corresponding to a high pilot spacing $\Delta_\text{P} \approx 2.5W_\text{coh}$.}
\label{fig:BER_SNR}
\end{figure}

\begin{figure}[!t]
\centering
\includegraphics[width=0.75\columnwidth]{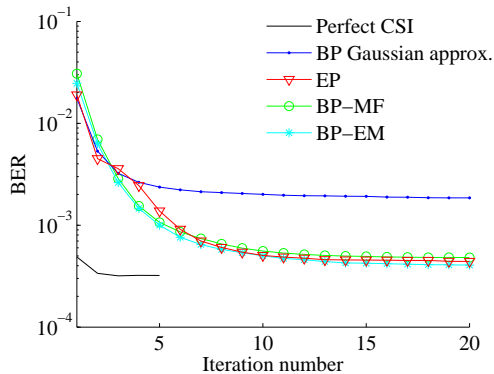}
\caption{Convergence of the BER performance from Fig.~\ref{fig:BER_SNR} at $\text{SNR}= 12\,\text{dB}$.}
\label{fig:BER_Iter}
\end{figure}

Results in terms of BER versus signal-to-noise ratio (SNR) are given in Fig.~\ref{fig:BER_SNR}, while the convergence of the BER with the number of iterations is illustrated in Fig.~\ref{fig:BER_Iter}.
The receivers based on EP, combined BP-MF and BP-EM exhibit similar performance. They significantly outperform the receiver employing BP with Gaussian approximation. Note that even with a high pilot spacing $\Delta_\text{P} \approx 2.5W_\text{coh}$ the performance of the former algorithms is close to that of the receiver having perfect CSI. These three algorithms converge in about 10--12 iterations, while BP with Gaussian approximation converges a little faster, but to a higher BER value. Other results not presented here show that for a higher pilot density the algorithms converge faster, as expected.

Note that the results for the (essentially equally-complex) BP-EM and BP-MF receivers are nearly identical, even if the former discards the soft information in channel estimation. We noticed during our evaluations that $\sigma^{2}_{h_i} \ll |\mu_{h_i}|^2$ even at low SNR values, so our explanation would be that accounting for $\sigma^{2}_{h_i}$ in the BP-MF receiver does not have a noticeable impact on the detection~\eqref{eq:m_fD_x_MF}.

\begin{table}[!h]
\caption{Parameters of the wireless OFDM system}
\label{tab:sim_param}
\begin{tabular}{l|l}
\hline
\textbf{Parameter} & \textbf{Value} \\
\hline \hline
Subcarrier spacing & $15\,\text{kHz}$ \\
Number of active subcarriers & $M + N = 300$ \\
Number of evenly spaced pilot symbols & $M = 10$\\
Pilot spacing & $\Delta_\text{P}\approx 500 \, \text{kHz}$ \\
Modulation scheme for data symbols & $16\, \text{QAM} \, (L=4)$\\
Convolutional channel code & $R = 1/3,\,(133,171,165)_8$ \\
Multipath channel model & 3GPP ETU \\
Coherence bandwidth of the channel & $W_\text{coh} \approx 200 \, \text{kHz}$\\
\hline
\end{tabular}

\end{table}


\section{Conclusions}

We formulated the problem of joint channel estimation and decoding in a communication system as inference in a graphical model.
To solve the inference problem, we resorted to a recently proposed message-passing framework that unifies the BP and MF algorithms and includes them as particular instances. Additionally, we illustrated how the combined framework can encompass the EP and EM inference algorithms.

Based on the inference techniques considered, we derived four
receiver algorithms. Since BP is not suitable for the studied
problem, as it leads to intractable messages, we applied its variant which employs Gaussian approximation of the computationally cumbersome messages instead. However,
our results showed that it performs significantly worse
than the other proposed schemes. Considering the BER
results, the computational complexity and stability of these
schemes, we conclude that the receiver based on the combined BP-MF
framework and its BP-EM variant are the most effective receiver
algorithms.

\section*{Acknowledgment}


Six projects have supported this work: the Project SIDOC under contract no. POSDRU/88/1.5/S/60078; the Cooperative Research Project 4GMCT funded by Intel Mobile Communications, Agilent Technologies, Aalborg University and the Danish National Advanced Technology Foundation; the PhD Project ``Iterative Information Processing for Wireless Receivers" funded by Renesas Mobile Corporation; the Project ICT-248894 WHERE2; the WWTF Grant ICT10-066; and the FWF Grant S10603-N13 within the National Research Network SISE.



%
\bibliographystyle{IEEEtran}
\bibliography{IEEEabrv,Infer_CE_Dec}

\end{document}